\documentclass[aps,prb,groupedaddress]{revtex4}

\usepackage{graphicx}% Include figure files
\usepackage{dcolumn}% Align table columns on decimal point
\usepackage{epsfig}
\usepackage{bm}% bold math
\usepackage{amsfonts}
\usepackage{latexsym}
\usepackage[dvipdfm]{hyperref}
\newcommand{\bfdelta}{ \mbox{\boldmath $\delta$} }
\newcommand{\sbfdelta}{\mbox{{\scriptsize\boldmath $\delta$}}}
\newcommand{\be}{\begin{equation}}
\newcommand{\ee}{\end{equation}}
\newcommand{\bea}{\begin{eqnarray}}
\newcommand{\eea}{\end{eqnarray}}

\newcommand{\half}{ \frac{1}{2} }
\newcommand{\bfsrho}{ \mbox{\scriptsize\boldmath $\rho$} }

\begin{document}

%\preprint{}
%\preprint{APS/123-QED}

\title{Further Series Studies of the Spin-$\frac{1}{2}$ Heisenberg Antiferromagnet at $T=0$:\\
Magnon Dispersion and Structure Factors}


% % \affiliation command applies to all authors since the last
% % \affiliation command. The \affiliation command should follow the
% % other information
\author{Weihong Zheng}
\email[]{w.zheng@unsw.edu.au}
\homepage[]{http://www.phys.unsw.edu.au/~zwh}
%\thanks{}
\affiliation{School of Physics,
The University of New South Wales,
Sydney, NSW 2052, Australia.}

\author{J. Oitmaa}
\email[]{j.oitmaa@unsw.edu.au}
%\thanks{}
\affiliation{School of Physics,
The University of New South Wales,
Sydney, NSW 2052, Australia.}

\author{C.J. Hamer}
\email[]{c.hamer@unsw.edu.au}
%\thanks{}
\affiliation{School of Physics,
The University of New South Wales,
Sydney, NSW 2052, Australia.}

%\author{R.R.P. Singh}
%\email[]{singh@raman.ucdavis.edu}
%%\homepage[]{Your web page}
%%\thanks{}
%%\altaffiliation{}
%\affiliation{Department of Physics, University of California, Davis,
%CA95616, USA}

%\date{Nov. 3, 1998}
\date{\today}

\begin{abstract}
We have extended our previous series studies of quantum antiferromagnets at zero
temperature by computing the one-magnon dispersion curves and various structure
factors for the linear chain, square and simple cubic lattices. Many of these
results are new; others are a substantial extension of previous work. These
results are directly comparable with neutron scattering experiments and
we make such comparisons where possible.
\end{abstract}

\pacs{PACS numbers:  75.10.Jm, 75.50.Ee, 75.40.Gb}

% 75.50.Ee Antiferromagnetics
% 71.10.Fd Lattice fermion models (Hubbard model, etc.)
% 71.27.+a Strongly correlated electron systems; heavy fermions
% 73.43.Nq Quantum phase transitions
% 75.40.Gb Dynamic properties (dynamic susceptibility, spin waves, spin diffusion, dynamic scaling, etc.)
% 75.10.Jm Quantized spin models
% 75.50.Gg Ferrimagnetics
% 75.10.Hk Classical spin models
% 74.70.Kn Organic superconductors
% 74.25.Dw Superconductivity phase diagrams
% 74.25.Bt Thermodynamic properties
% 74.25.Ha Magnetic properties
% 74.25.Kc Phonons
% 76.50.+g Ferromagnetic, antiferromagnetic, and ferrimagnetic
% 71.10.Li Excited states and pairing interactions in model systems

\keywords{Heisenberg model, antiferromagnets, series expansions, magnon dispersion, structure
factor, neutron scattering}

\maketitle

\section{\label{sec:intro}INTRODUCTION}

The spin-$\half$ Heisenberg antiferromagnet, which we take in the exchange anisotropic
form
\be
H = J \sum_{\langle ij\rangle } [ S_i^z S_j^z + \lambda ( S_i^x S_j^x + S_i^y S_j^y ) ]~,  \quad | \lambda | \leq 1
\label{eq_H}
\ee
is the archetypal model for describing long-range antiferromagnetic order in solids.
Although there are no exact solutions in greater than one spatial dimension,
a great deal is known about the model from various systematic approaches: exact
diagonalizations, quantum Monte Carlo methods, and series expansions. Good overviews of
the subject, with a particular focus on the square lattice and the relation to the
high T$_c$ cuprate superconductors, have been given by Barnes\cite{Barnes91} and
Manousakis\cite{Manousakis91}. An area of particular current interest is the relation of models
such as (\ref{eq_H}) to real materials. Quantities that can be most readily compared
are the dispersion relations of low energy quasiparticle
excitations and dynamical or integrated structure factors . The calculation of these is the main thrust of the current paper.
At the same time the building of new and more powerful neutron scattering facilities
is providing more precise data and allowing more detailed comparisons between
experiment and theory\cite{Ronnow,Kim,Radu}.

Our approach is through high-order `linked cluster' series expansions\cite{gel00},
where the quantities of interest are expanded perturbatively in powers of $\lambda$
(the so-called {\it Ising} expansion), and numerically evaluated at $\lambda=1$.
This approach has been used with considerable success in computing ground state properties
of quantum antiferromagnets\cite{rajiv89,zheng_gs,oit94}, and in computing the magnon excitation spectrum
and spectral weight for the square lattice\cite{rajiv}.
In our calculations we set $J=1$ to determine the energy scale, except in comparison with
experiment.
In Section \ref{sec2} we will define the various quantities of interest, and give a brief
overview of the methodology. Section \ref{sec3} gives new results for the
structure factors for the linear chain. % and explains a puzzle in earlier work\cite{singh89}.
Section \ref{sec4} extends previous work for the square lattice\cite{zheng_gs,rajiv} and
gives new results for the longitudinal and total structure factors. Section \ref{sec5}
gives results for the simple cubic lattice. Ground state series are extended by 2 terms and
series results for the magnon energies and all structure factors are given for the first time.
Finally in Section \ref{sec6} we summarize and attempt to  relate our work to experiment.

\section{\label{sec2}Methodology and Definitions}
The essence of the linked cluster method\cite{gel00} is the realization
that many properties of a lattice model, in the thermodynamic limit
$N\to \infty$, can be expressed as a sum of contributions from all possible
connected or linked clusters of sites which can be embedded in the
particular lattice of interest. This is most obvious in the case of
extensive bulk properties, such as the ground state energy, magnetization,
susceptibility, etc, where we have
\be
F_N (x ) = \sum_{\{g\} } C(g/{\cal L}) f_g (x )
\ee
where $F_N(x )$ is the quantity of interest, with $x $ representing
the set of parameters in the Hamiltonian. The sum is over all clusters
$\{ g \} $, with $C(g/{\cal L})$ being the embedding constant of cluster $g$
in the lattice ${\cal L}$ of  $N$ sites (proportional to $N$) and
$f_g (x)$ a {\it reduced} quantity for cluster $g$. These latter quantities,
which are independent of the lattice, are computed recursively\cite{gel00}.
It is easy to show that $f_g(x)$ is zero for any disconnected cluster,
provided $F$ is an extensive quantity. Linked cluster series expansions are
then obtained by writing the Hamiltonian in the usual form for perturbation theory,
$H=H_0 + \lambda V$, and calculating the cluster contributions perturbatively,
as series in $\lambda$, up to some maximum achievable  order (typically 10-20).
The bulk series for $F_N(\lambda)$ is then evaluated at fixed $\lambda$, or
extrapolated to $\lambda=1$, via standard numerical methods such as Pad\'e
approximants or integrated differential approximants\cite{gut}. In practice
all of this is done by computer and it is feasible to deal with of order $10^6$
distinct clusters.
%In this calculation, we also add a staggered field
%\be
%h (1-\lambda) \sum_i (-1)^i S_i^z
%\ee
%to the Eq. (\ref{eq_H}), and adjust $h$ to get best convergence series.

%\be
%H = J \sum_{\langle ij\rangle } [ - S_i^z S_j^z + \lambda ( S_i^x S_j^x - S_i^y S_j^y ) ]  \quad | \lambda | \leq 1
%\label{eq_H2}
%\ee

A  stringent comparison between real materials and theoretical models is often provided
by the spectrum of low energy excitations. These excitation energies can be measured
in scattering experiments, and are characteristic of the quantum dynamics
of the system. Gelfand\cite{gel96} first showed how to compute excitation energies
perturbatively, within a linked-cluster approach, and this is now a standard technique\cite{gel00}.
The basic idea is to compute an {\it effective Hamiltonian} matrix, which operates
in the subspace of  one-particle excitations of a cluster, use this to
obtain a set of {\it transition amplitudes} $t({\bf r})$ which describe propagation of the
excitation through a distance ${\bf r}$, obtain transition amplitudes for the bulk
lattice by summing over clusters, and finally take the Fourier transform, giving the
excitation energy in ${\bf k}$ space
\be
\epsilon ({\bf k}) = \sum_{{\bf r}} t({\bf r}) e^{i {\bf k} \cdot {\bf r}} \label{eq_mk}
\ee
While the dispersion relation (\ref{eq_mk}) is an important probe of the quantum
dynamics, an even more comprehensive probe is the dynamical structure factor
\be
S_{\alpha} ({\bf k}, \omega ) = {1\over 2 \pi} \int_{-\infty}^{\infty} dt e^{i \omega t} \sum_{\bf r}
e^{i {\bf k}\cdot {\bf r}} \langle S_0^{\alpha} (0) S_{\bf r}^{\alpha} (t) \rangle_0
\ee
i.e. the spatial and temporal Fourier transform of the dynamical spin-spin correlation function.
 The angular brackets denote an average (here a ground state expectation value), and $\alpha=x,y,z$.
 This quantity is directly related to the cross section for inelastic neutron scattering (see e.g.
Broholm \& Aeppli \cite{Broholm}). The integrated or static structure factor
\bea
S_{\alpha} ({\bf k}) &=& \int_{-\infty}^{\infty} d\omega S_{\alpha} ( {\bf k}, \omega) \nonumber \\
&=& \sum_{\bf r} e^{i {\bf k}\cdot {\bf r}} \langle S_0^{\alpha} S_{\bf r}^{\alpha} \rangle
\eea
is measured in an experiment where all neutron energies are included.

For an isotropic system, in the absence of long-range magnetic order or other spontaneously broken
symmetry, the components $\alpha=x,y,z$ of $S_{\alpha} ({\bf k}, \omega) $ or
$S_{\alpha} ({\bf k} ) $ will be equal. This will no longer be the case if magnetic order is present.
For a collinear ordered state, in the $z$ direction, we need to distinguish between a
{\it longitudinal} structure factor
\be
S_l ({\bf k}) = \sum_{\bf r} e^{i {\bf k}\cdot {\bf r}} [ \langle S_0^z S_{\bf r}^z \rangle
-  \langle S_0^z  \rangle \langle  S_{\bf r}^z \rangle ] \label{eq_lsf}
\ee
and a {\it transverse} structure factor
\be
S_t ({\bf k}) = \sum_{\bf r} e^{i {\bf k}\cdot {\bf r}} \langle S_0^x S_{\bf r}^x +
 S_0^y S_{\bf r}^y \rangle  \label{eq_tsf}
\ee
If unpolarized neutrons are used the cross section will measure the total structure factor
\be
S_{\rm tot} ({\bf k}) = S_l ({\bf k}) + S_t ({\bf k})
\ee
The dominant contribution to the transverse dynamical structure factor will come from
one-magnon excitations, and $S_t ({\bf k}, \omega )$ will have the form
\be
S_t ({\bf k}, \omega ) = A_1 ({\bf k}) \delta (\omega - \epsilon ({\bf k})) + S_{\rm inc} ({\bf k}, \omega)
\ee
where $A_1 ({\bf k})$ is called the one-magnon spectral weight (or the exclusive structure factor) and
$ S_{\rm inc} ({\bf k}, \omega)$ is a smooth incoherent background term,
arising from multi-magnon processes. It is easy to show that
\be
A_1 ({\bf k}) = \half \sum_{\bf r} e^{i {\bf k}\cdot {\bf r}}
\langle \Psi_0       \vert ( S_0^+ + S_0^-)              \vert \Psi_{\bf k} \rangle
\langle \Psi_{\bf k} \vert ( S_{\bf r}^+ + S_{\bf r}^- ) \vert \Psi_0 \rangle  \label{eq_S1p}
\ee
where $\vert \Psi_0 \rangle$, $\vert \Psi_{\bf k} \rangle$ are respectively the
ground state and one-magnon state and  $S_{\bf r}^+$, $S_{0}^-$ are spin raising
and lowering operators. It is also useful to define a relative multi-magnon
spectral weight by
\be
W_t ({\bf k}) = 1- A_1 ({\bf k})/S_t ({\bf k}) \label{eq_W_t}
\ee
and a similar quantity for unpolarized neutron scattering
\be
W_{\rm tot} ({\bf k}) = 1- A_1 ({\bf k})/S_{\rm tot} ({\bf k}) \label{eq_W_tot}
\ee

The linked cluster formalism to compute the structure factor is relatively straightforward,
and has been discussed in Refs. \onlinecite{zwh_weight} and \onlinecite{rajiv}. The correlator sums
\be
Z_{\alpha} ( {\bf r}) \equiv  \sum_i \langle S_i^{\alpha} S_{i+{\bf r}}^{\alpha} \rangle
\ee
are extensive quantities and thus have a linked-cluster expansion.
There is, however, one interesting and important
point regarding  the longitudinal correlators and the structure factor.
Linked cluster series for the correlators
$\langle S_0^{z} S_{{\bf r}}^{z} \rangle$,
computed from a set of clusters up to some fixed maximum size, will have a
maximum order in $\lambda$ which decreases with increasing ${\bf r}$. On
the other hand, the series for the {\it compensated} correlator
$\langle S_0^{z} S_{\bf r}^{z} \rangle - \langle S_0^{z} \rangle \langle S_{\bf r}^{z} \rangle $
has a maximum order independent of ${\bf r}$.
%because they can be expressed purely in terms of connected diagrams.(?)
This can be understood as follows. For any cluster the longitudinal correlator series
all start with a constant ($\lambda^0$) term. Subtraction of subgraph contributions
will cause cancellation of leading terms, leaving a series starting with some minimum power $\lambda^{p_{\rm min}}$.
However $p_{\rm min}$ decreases with increasing $r$, and is zero for $r=r_{\rm max}$,
the largest correlator which fits into the cluster, since, in this case, there are no
subgraph subtractions. Thus, in the absence of the compensating term (\ref{eq_add_term_lsf}),
much larger clusters would be required to give the large-$r$
correlator series to the  same order. Inclusion of the compensating term
avoids this problem since the leading terms in the bare correlator cancel and
$p_{\rm min}$ (defined above), after subgraph subtraction, is independent of $r$.
This allows longer series to be
derived for the structure factor as defined in
(\ref{eq_lsf}).
% in fact, it make the calculations possible by using the series expansions.
The additional term
\be
\sum_{\bf r} e^{i {\bf k} \cdot {\bf r}} \langle S_0^z \rangle \langle S_{\bf r}^z \rangle \label{eq_add_term_lsf}
\ee
will give a delta function peak at the antiferromagnetic wavevector ${\bf k}_{\rm AF}$,
but will not change the longitudinal structure factor for ${\bf k}\neq {\bf k}_{\rm AF}$.
The inclusion of this term reduces the total longitudinal structure factor,
summed over momentum ${\bf k}$,
% the autocorrelation function (i.e. the integral over momentum) for the longitudinal structure factor
% the momentum integral of the longitudinal structure factor
% ,i.e. the area of longitudinal structure factor,
from $S^2$ to $S^2-M^2$, where $S$ and $M$ are the spin and staggered magnetization, respectively.
For the transverse structure factor this total sum is just $S$.
%Note that the autocorrelation function for the transverse structure factor is $S$.

There are two methods for computing series for the one-magnon spectral weight $A_1({\bf k})$.
The first is to proceed directly from Eq. (\ref{eq_S1p}), as in Ref. \onlinecite{rajiv}.
An alternative method\cite{zwh_weight} is from the linked cluster series for another
 quantity, the so-called `exclusive matrix element',
\be
\Omega (\bfdelta) = \langle \Psi_0  \vert ( S_i^+ + S_i^-)   \vert \Psi_{m} \rangle~; \quad \bfdelta = {\bf r}_i - {\bf r}_m
\ee
where $\vert \Psi_{m} \rangle$ is the one-magnon wavefunction with initial unperturbed
excitation at site $m$. Then
\be
A_{1} ({\bf k}) = {\Big \vert} \sum_{\sbfdelta} \Omega (\bfdelta) e^{i {\bf k}\cdot \sbfdelta} {\Big \vert}^2
\ee
The advantage of this second method is that it can be easily extended to the two-particle case, although
we do not pursue this here. The two methods should, of course, result in the same final series.
This provides a useful check on the correctness of the input cluster data, more
stringent than the calculation of ground state bulk properties or excitation spectra.

% %Note: We can get the series for correlation length from integrated structure factor, do we want to mention this ???
% From the series of integrated total structure factor, one can also obtain the series for correlation
% length $\xi$ via
% \be
% \xi^2 = { \sum_{ij} r_{ij}^2 \pi_{ij} \langle {\bf S}_i \cdot {\bf S}_j \rangle_0 \over
%  \sum_{ij} \pi_{ij} \langle {\bf S}_i \cdot {\bf S}_j \rangle_0 }
% \ee
% where $r_{ij} $ is the Euclidean distance between spins $i$ and $j$, and the parity factor $\pi_{ij}$
% takes values $\pm 1$, depending on whether or not spin $i$ lies in the same sublattices as spin $j$.
% (???, maybe we can not get $\xi$, due to the extra term in Eq. {\ref{eq_add_term_lsf}.
% Ignoring this extra term, the critical index I got for $\xi^2$ for square lattice is -1.15,
% rather than -1.)

We compare our series results with the prediction from spin-wave calculations.
For the anisotropic Hamiltonian  (\ref{eq_H}),
the spin-wave theory has been computed to 4th order for the ground state energy, and 3rd order
for most other properties\cite{zwh_sw}.
The second order spin-wave theory predicts the spin-wave excitation spectrum
\be
\epsilon_{\rm k} = z S q_{\bf k} - {z\over 2}
{\big [} C_{-1} q_{\bf k} + (\lambda^{-2}-1) (C_{-1}-C_1) (q_{\bf k}^{-1} - q_{\bf k}) {\big ]}
\ee
where $z$ is the lattice coordination number,
$q_{\bf k} = (1-\lambda^2 \gamma_{\bf k}^2)^{1/2}$,
$C_n$ is defined as
\be
C_n = {2\over N} \sum_{\bf k} [ (1-\lambda^2 \gamma_{\bf k}^2)^{n/2}-1]
\ee
and %the lattice structure factor $\gamma_{\bf k}$ is
\be
\gamma_{\bf k} = {1\over z} \sum_{\bfsrho} e^{i {\bf k}\cdot \bfsrho}
\ee
At $\lambda=1$, we can get a simple expression for the excitation spectrum
\be
\epsilon_{\rm k} = z S (1- \gamma_{\bf k}^2)^{1/2} [ 1-C_{-1}/(2S) ]
\ee
That is, the second order spin-wave theory only gives an overall a renormalization,
with renormalization factor $Z_c=1-C_{-1}/(2S)$, to the dispersion given
by linear spin-wave theory.

Linear spin-wave theory
gives the transverse structure factor as
\be
S_{\rm t} ({\bf k}) = S \sqrt{ {1-\lambda \gamma_{\bf k} \over 1 + \lambda \gamma_{\bf k}} }
\ee
In the limit $ k = |{\bf k}| \to 0$, $\gamma_{\bf k}  \to 1 - k^2/z$,
\be
S_{\rm t} ({\bf k}) =
    S \left( {1-\lambda \over 1+\lambda} + k^2 {2 \lambda \over (1+\lambda)^2 z} \right)^{1/2}
\ee
so $S_{\rm t} ({\bf k})$ vanishes as $Sk/\sqrt{2z}$ at $\lambda=1$,
while at ${\bf k}=0$, $S_{\rm t}$ vanishes as $S(1-\lambda)^{1/2}/\sqrt{2}$ as $\lambda\to 1$.

In the limit $ q \equiv |{\bf q}|  = |{\bf k}_{\rm AF} - {\bf k}| \to 0$,
$\gamma_{\bf k}  \to -1 + q^2/z$, and
\be
1/S_{\rm t} ({\bf k}) =
    S^{-1} \left( {1-\lambda \over 1+\lambda} + q^2 {2 \lambda \over (1+\lambda)^2 z} \right)^{1/2} \label{St_sw_asym}
\ee
so $S_{\rm t} ({\bf k})$ diverges as $S \sqrt{2z}/q$ at $\lambda=1$,
while at ${\bf k}={\bf k}_{\rm AF}$, $S_{\rm t}$ diverges as $S\sqrt{2}(1-\lambda)^{-1/2}$ as $\lambda\to 1$.

%At the limit ${\bf k}\to 0$, $\gamma_{\bf k}  \to 1 - {\bf k}^2/z$,
%while at limit ${\bf q} = {\bf k}_{\rm AF} - {\bf k} \to 0$,
%$\gamma_{\bf k}  \to -1 + {\bf q}^2/z$, this implit that for $\lambda=1$
%\be
%S_{\rm t} ({\bf k}) =
%\left\{%
%\begin{array}{ll}
%     {S\over \sqrt{2z}} k, & \hbox{ $k = |{\bf k}| \to 0$ ;} \\
%    {S \sqrt{2z} \over q}, & \hbox{  $q =|{\bf q}| \to 0$ .} \\
%\end{array}%
%\right.
%\ee
%or as $\lambda\to 1$,
%\be
%S_{\rm t} ({\bf k}) =
%\left\{%
%\begin{array}{ll}
%    {S\over \sqrt{2}} (1-\lambda)^{1/2} , & \hbox{ ${\bf k} =  0$ ;} \\
%    {\sqrt{2} S  (1-\lambda)^{-1/2}}, & \hbox{${\bf k} ={\bf k}_{\rm AF}$ .} \\
%\end{array}%
%\right.
%\ee
%

We now turn to the series results.

\section{\label{sec3}The Linear Chain}
The anisotropic spin-$\half$ Heisenberg  antiferromagnet in one dimension
(the XXZ chain) has been the subject of much study. Many materials which
are well represented by this model have been identified (see Table 1 in Ref.
\onlinecite{Broholm2}). The possibility of exact results via Bethe ansatz methods
has led to a good overall theoretical understanding of the
model. In particular it is known that the elementary excitations
are $S=\half$ spinons, or domain walls, with a dispersion relation\cite{exact}
\be
\epsilon_{\rm spinon} (k) = I [ \cos^2 (k) + g^2 \sin^2 (k) ]^{1/2} \label{mk_1D_exact}
\ee
where
\be
I = (1-\lambda^2)^{1/2} K(g'^2)/\pi~, \quad g'^2 = 1 -g^2  \label{eq_I_1D}
\ee
and $g$ is the solution of
\be
\pi K(g^2)/K(g'^2) = {\rm sech}^{-1} (\lambda)  \label{eq_KK_1D}
\ee
and $K$ denotes the complete elliptic integral,
\be
K(x) = \int_0^{\pi/2} [ 1 - x \sin^2 (\theta)]^{-1/2} d \theta
\ee

A series expansion for the spinon energy has already been derived by Singh\cite{ragiv_1D},
and shown to agree precisely with the expansion of the exact result (\ref{mk_1D_exact})
in powers of $\lambda$.

\begin{figure}[htb]
\centering
      \includegraphics[width=0.45\textwidth]{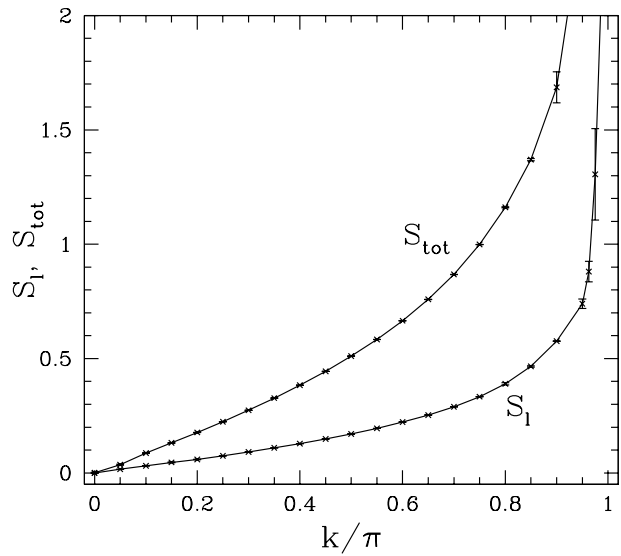}
      \caption{The total and longitudinal structure factor for the linear chain.\label{fig_sf_1D}}
\end{figure}

The  structure factors are not known exactly for the XXZ chain, and here series expansions have
a role to play. Singh {\it et al.}\cite{singh89} obtained long series for the longitudinal and transverse structure
factors (\ref{eq_lsf}) and (\ref{eq_tsf}) at the antiferromagnetic wavevector $k=\pi$, to 22 and 12 terms respectively
in $\lambda$ (only even terms occur in the longitudinal case) and
studied the divergence of both quantities as $\lambda\to 1-$. They found different exponents $(\sim 1.0, 0.75)$
for the two power laws, and explained this apparently surprising result via a renormalization group argument.

We have computed series for all of the structure factors, for general wavevector $k$, to order $\lambda^{28}$.
This represents 16 additional terms for the transverse (and hence the total) structure factor.
Our results for the isotropic case ($\lambda=1$) are shown in Figure \ref{fig_sf_1D}.
The structure factors diverge at $k=\pi$, as expected. For $k\neq \pi$,
we find, to numerical accuracy, that $S_{\rm tot} = 3 S_{\rm l}$,
% Fig. \ref{fig_sf_1D} shows our results at $\lambda=1$, we can see that all tructure factors diverge at $k=\pi$,
% for $k\neq \pi$, the total structure factor is three times of longitudinal structure factor,
as expected, since the system
has no long range order. For $k=\pi$, our longer series also show that
longitudinal and transverse structure factors diverge with two different  exponents,
 as found by Singh {\it et al.}\cite{singh89}.
%  We believe the occurrence of the  two different exponents
% is due to the additional term, Eq.(\ref{eq_add_term_lsf}), added to the longitudinal structure factor.
%%For small $k$, $S_{\rm l}(k) \propto k^2$, while $S_{\rm t}(k) \propto k$.

\section{\label{sec4}The Square Lattice}
The square lattice $S=\half$ antiferromagnet has been much studied in recent
years, largely due to its relevance to the high $T_c$ cuprate superconductors.
There is convincing, though not yet rigorous, evidence that the ground state
has long-range N\'eel order, reduced by quantum fluctuations.

\begin{figure}[htb]
\centering
      \includegraphics[width=0.45\textwidth]{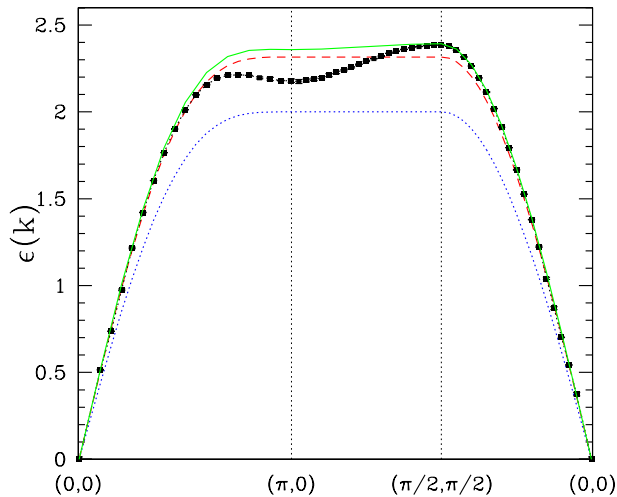}
      \caption{(Color online) The 1-magnon excitation spectrum
$\epsilon ({\bf k})$ along high-symmetry cuts through the Brillouin
zone for the Heisenberg antiferromagnet on a square lattice. Also shown are
      the results of  first order (blue dotted line), second order (red dashed line)
      and third order (green solid line) spin-wave theory.\label{fig_mk_sq}}
\end{figure}

Some years ago we derived\cite{zheng_gs} perturbation series for the ground state energy,
sublattice magnetization and parallel susceptibility to 14th order in the exchange
anisotropic parameter $\lambda$, and for the transverse (perpendicular) susceptibility
to order 13. These series provided very precise estimates of ground state properties
for the entire range $0<\lambda\leq 1$, including the isotropic point $\lambda=1$.
We also showed that higher order spin-wave theory\cite{zwh_sw} was in excellent agreement
with the series results. We have recently extended these series by two terms,
to order $\lambda^{16}$, the calculation involving a list of 185\,690 clusters, up to 16 sites.
 We are happy to provide the new coefficients on request,
but do not present any new analysis of ground state properties here.

\begin{figure}[htb]
\centering
      \includegraphics[width=0.45\textwidth]{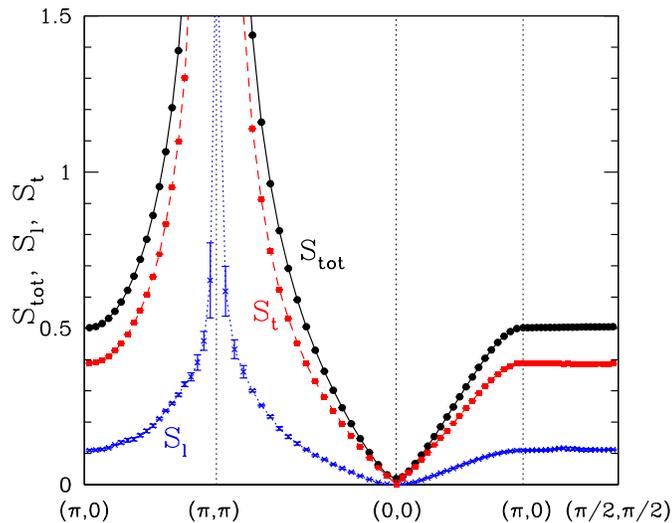}
      \caption{(Color online) The various integrated structure factors $S_{\rm tot}$ (unpolarized),
       $S_{\rm t}$ (transverse)  and $S_{\rm l}$ (longitudinal) along high-symmetry cuts through the Brillouin
zone for the Heisenberg antiferromagnet on a square lattice.\label{fig_tot_wei_sq}}
\end{figure}

We have also extended an earlier calculation \cite{rajiv} of the magnon excitation
spectrum and spectral weight series by four terms, to order $\lambda^{14}$.
This calculation involves a large list of 4\,654\,284 clusters, up to 15 sites.
% which take 10GB disk space.
The series coefficients are quite extensive
and are not presented here, but we will provide them on request. We give
in Table \ref{tab_sq} the series at ${\bf k} = (\pi,\pi)$, $(\pi,0)$, and $(\pi/2,\pi/2)$.
The resulting
magnon dispersion curve %, obtained by extrapolating the series to $\lambda=1$,
is shown in Fig. \ref{fig_mk_sq}.
It was obtained by extrapolating the series to $\lambda=1$, using integrated
differential approximants.
 The first, second and third order spin-wave results\cite{zheng_gs,zwh_sw}
are included for comparison. We confirm the overall shape of the dispersion curve
obtained previously\cite{rajiv} but provide greater precision from the longer
series. It is evident from the figure that the dispersion curve along the edge of the
magnetic Brillouin zone $(\pi, 0)\to (\pi/2, \pi/2)$ is not flat, as
predicted by the first and second order spin-wave theory. We find numerically
% we get the energies at $(\pi,0)$ and $(\pi/2, \pi/2)$
\be
\epsilon(\pi,0)=2.18(1)~, \quad \epsilon(\pi/2,\pi/2)=2.385(1)
\ee
and so there is a 9.4\%
increase from $(\pi,0)$ to $(\pi/2,\pi/2)$.
This agrees very well with a recent  quantum Monte Carlo calculation\cite{sandvik}
$\epsilon(\pi,0)=2.16$, $\epsilon(\pi/2,\pi/2)=2.39$.
Spin-wave theory, however, is unable to reproduce this variation even at third
order\cite{zwh_sw} (via both Holstein-Primakoff and Dyson-Maleev transformations), 
which  gives
$\epsilon(\pi,0)=2.35858$, $\epsilon(\pi/2,\pi/2)=2.39199$.
Our series results are also in qualitative agreement
with experimental data for
 Cu(DCOO)$_2\cdot 4$D$_2$O (CFTD)\cite{Ronnow} and
Sr$_2$Cu$_3$O$_4$Cl$_2$\cite{Kim}.
However in La$_2$CuO$_4$ the observed magnon energy
at $(\pi,0)$ is higher than at $(\pi/2,\pi/2)$\cite{Radu},
opposite to the model result. It has been suggested\cite{Radu} that this is due to the presence
of a significant ring exchange term in this material,
but other explanations are possible\cite{hubbard}.
% alternatively one can explain
% this nicely via hubbard model\cite{??}.

\begin{figure}[htb]
\centering
      \includegraphics[width=0.45\textwidth]{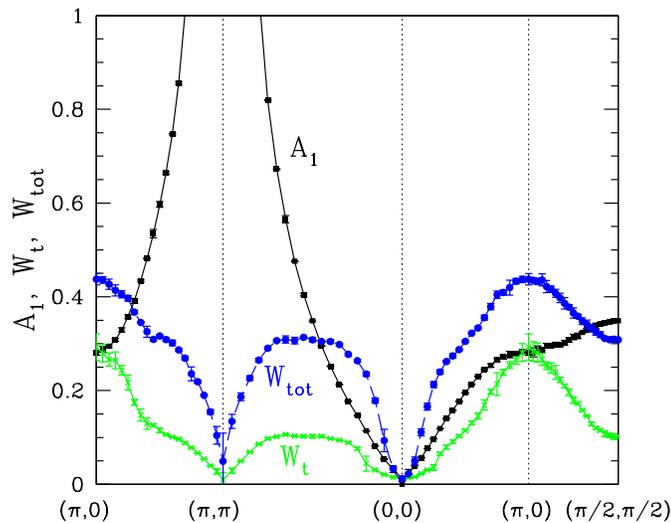}
      \caption{(Color online) The 1-magnon spectral weight $A_{\rm 1}$,
      and  multi-magnon spectral weights
      $W_{\rm t}$ and $W_{\rm tot}$ for the Heisenberg antiferromagnet on a square lattice.\label{fig_wei_1p_sq}}
\end{figure}

From our series for the magnon energies we can obtain a rather precise
estimate of the spin wave velocity $v$. Following Singh and Gelfand\cite{rajiv}
we write the magnon energy at long wavelength in the form
\be
\epsilon ({\bf k}) = C(\lambda) + D(\lambda ) k^2 + O({\bf k}^3)~, \quad k = \vert {\bf k} \vert \to 0 \label{eq_v1}
\ee
% The series for $C(\lambda)$ is the series of minimum energy gap,
% the series for $D(\lambda)$ is also given in Table \ref{tab_sq}.
% At the limit $\lambda \to 1$, $C\to 0$, $D\to \infty$. We known at ${\bf k}\to 0$,
% $\epsilon ({\bf k}) $ have the form,
% \be
% \epsilon ({\bf k}) = (A(\lambda) + B(\lambda ) k^2)^{1/2}  \label{eq_v2}
% \ee
% and at limit $\lambda \to 1$, $A\to 0$, $B\to v^2$, compare  Eq. \ref{eq_v1} and \ref{eq_v2},
% one can get $A=C^2$ and $B=2 CD$,  so
The spin wave velocity $v^2$ can be obtained from
the series for $2 C(\lambda) D(\lambda )$,
evaluated at $\lambda=1$.
Using  integrated differential approximants\cite{gut}, we
estimate $2CD=2.774(6)$ at $\lambda=1$,
and conclude that
$v/Ja =1.666(2)$.
For ${\bf k}=0$, we expect the spin-wave energy to vanish as
\be
\epsilon ({\bf k}=0) = c (1-\lambda^2)^{1/2}, \quad \lambda \to 1- \label{eq_gap_k0}
\ee
where the coefficient $c$ can be estimated from our series: the result is $c=1.256(2)$.
%we can also get the gap vanish as $\epsilon ({\bf k}=0) = 1.256(2) (1-\lambda^2)^(1/2)$.
Third order spin-wave theory\cite{zwh_sw} gives
$v/Ja = 1.66802$ and $c=1.23531$, agreeing with the series estimates within 2\%.

\begin{figure}[htb]
\centering
      \includegraphics[width=0.45\textwidth]{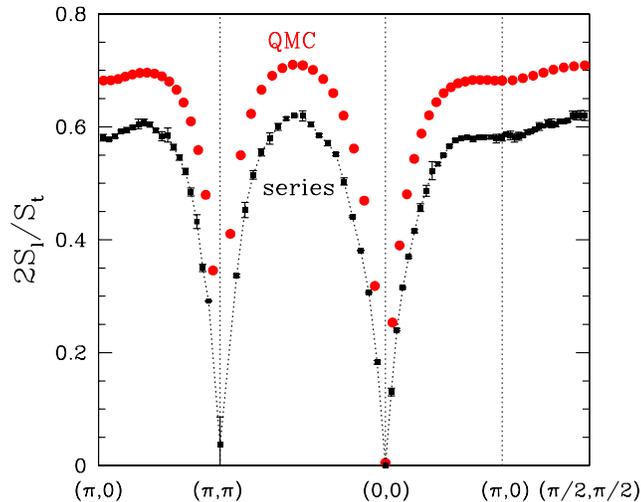}
      \caption{(Color online) Ratio of longitudinal and transverse structure factors $2S_{\rm l}/S_{\rm t}$
      for the Heisenberg antiferromagnet on a square lattice. Also shown, for comparison,
       is the QMC results\cite{sandvik}.
       % with the point at ${\bf k}=(\pi,\pi)$ removed.
       \label{fig_ratio_Szz_Sxy_sq}}
\end{figure}

In Figure \ref{fig_tot_wei_sq} we show results for the various integrated structure factors along high
symmetry lines in the Brillouin zone.
The transverse structure factor was computed previously\cite{rajiv} to order $\lambda^{10}$ -
we have extended this series by four terms, to order $\lambda^{14}$. Calculation of the
longitudinal structure factor, and hence the total structure factor, by series methods is,
as far as we know, given here for the first time. The series at ${\bf k}=(\pi,\pi)$,
$(\pi,0)$, and $(\pi/2,\pi/2)$
are listed in Table \ref{tab_sq}.
Various features deserve comment. Both longitudinal
and transverse structure factors vanish at ${\bf k}=(0,0)$. It is known, on general
grounds, that the $k$ dependence at this point is $k^2$, $k$ respectively. Hence the
longitudinal structure factor vanishes more rapidly.
We estimate, from our series,
%We can easily get the coefficient of $k^2$ for $S_l$ (the series for the coefficient of $k^2$
%is given in Table \ref{tab_sq}), since $S_l$ is zero at ${\bf k}=0$, the
%results are
\bea
S_l ({\bf k}) &=& 0.042(4)k^2  \quad {\rm as} \quad k = \vert {\bf k}\vert  \to 0 \\  %  2D=  8.417810D-02  6.2D-03
S_t ({\bf k}) &=& 0.108(4) k  \quad {\rm as} \quad k = \vert {\bf k}\vert  \to 0    %v^2= 1.169223D-02  7.3D-04 (d1)
\eea
where the coefficient for $S_t ({\bf k})$ is estimated using the same method as used for the spin-wave velocity $v$.
A second-order spin-wave calculation\cite{canali} gives
 $S_{\rm t} ({\bf k})=0.10133 k$.

%To get the coefficient of $k$ at the limit $k\to 0$ for $S_t$ and $S_{\rm tot}$, one can use  exactly
%the same trick as the spin-wave velocity $v$ mentioned above, and the results are
%\bea \label{eq_S_k0_sq}
%S_{\rm tot}(k)&=& 0.109(2) k \quad {\rm as} \quad k = \vert {\bf k}\vert  \to 0 \\ %v^2= 1.191005D-02  3.3D-04 (d1)
%S_t ({\bf k}) &=& 0.108(4) k \quad {\rm as} \quad k = \vert {\bf k}\vert  \to 0    %v^2= 1.169223D-02  7.3D-04 (d1)
%\eea

Both structure factors diverge at
the antiferromagnetic wave vector ${\bf k}=(\pi,\pi)$. If the N\'eel state were
an exact eigenstate the static longitudinal structure factor would be zero,
except for a $\delta$-function
peak at $(\pi,\pi)$. The actual shape reflects the additional contribution from quantum
fluctuations.
We first consider the asymptotic behaviour of longitudinal and transverse static structure
factors at ${\bf k}=(\pi,\pi)$, as $\lambda\to 1$. Assuming
% Before we discuss the critical behaviour near ${\bf k}=(\pi,\pi)$ for
% $\lambda=1$, let us discuss the critical behaviour near $\lambda=1$ for ${\bf k}=(\pi,\pi)$
\be
S_{\rm l}(\lambda) \sim (1-\lambda)^{-\sigma_l}~, \quad
S_{\rm t}(\lambda) \sim (1-\lambda)^{-\sigma_t}
\ee
we estimate, from biased Dlog Pad\'e approximants, that
% As the linear chain case, the biased Dlog Pad\'e approximant to $S_l$ and $S_{t}$ (and so $S_{\rm tot}$)
% show their critical index at $\lambda =1$ is different,
% the critical index for $S_t$ is clearly
$\sigma_t =0.50(2)$,
while  $\sigma_l = 0.3(1)$.
The exponents again differ, as in the 1D case,
but here it is $\sigma_l$ which is apparently smaller (this could be related
to the fact that $\langle S^z \rangle \neq 0$ on the square lattice).
Linear spin-wave theory gives $\sigma_t =1/2$ (see Eq. (\ref{St_sw_asym})),
but one would need a higher-order calculation to give
$\sigma_l$, which has not yet been done.
Next we consider the way in which the transverse and total structure factors
at $\lambda=1$ diverge as ${\bf k}\to (\pi, \pi )$. Defining
${\bf q} = (\pi,\pi)-{\bf k}$, we write
% which turn out
%to be smaller than $\sigma_t$, (this is opposit to the case for linear chain).
%From this one expect that $S_{\rm t}$ and $S_{\rm tot}$ at $\lambda=1$ should diverge
%as $q^{-1}$  with $q = \vert (\pi,\pi)-{\bf k} \vert$, to get their coefficient,
%one can use a similar trick as the spin-wave velocity, i.e. we expand the
%series $S_{\rm t}$ (or $S_{\rm tot}$) near $q=0$
\be
S_{\rm t} ({\bf q}) = C(\lambda) + D(\lambda ) q^2 + O({\bf q^3})~,
\quad q = \vert {\bf q} \vert  \to 0
\ee
Both $C(\lambda)$ and $D(\lambda)$ diverge at $\lambda = 1$. However if we compute the inverse
\be
1/S_{\rm t} ({\bf q}) = 1/C(\lambda) - D(\lambda ) q^2/C^2(\lambda) + O({\bf q}^3)
\ee
and compare with the  asymptotic form (see Eq. \ref{St_sw_asym})
\be
1/S_{\rm t} ({\bf q}) = [ A (\lambda) + B(\lambda ) q^2]^{1/2}
\ee
we find that $S_{\rm t}$ diverges as $(B^{1/2}q)^{-1}$ with
$B=-2D/C^3$.
The series for $D$ for $S_{\rm t}$ is given in Table \ref{tab_sq}.
Our series, when analysed in this way, gives
\be  % 1.55827046D-02  4.0D-04 (t=0)    1.59952368D-02  2.5D-04  (t=1)
S_{\rm t} (q) = 0.93(7)/q , \quad q\to 0 \\  %v^(-2)=1.15(15) (d1)  1.05(8)(x1)
%S_{\rm tot}(q)&=& 0.95(5)/q , \quad q\to 0   %v^(-2)=1.1(1)
%A_{\rm 1} (q) &=& 0.97(5)/q , \quad q\to (\pi,\pi,\pi) \\  %v^(-2)=1.05(10)(d1)   1.051D+00  5.5D-03
\ee
The total structure factor series gives an estimate of $0.95(5)$,
consistent with the same result.
%The above results at the limit $k\to 0$ and $q\to 0$
%agree very well with the results $S_{\rm t} ({\bf k})=0.10133 k$,
% $S_{\rm t}({\bf q})=0.9288/q$
%obtained by the
Spin-wave theory\cite{canali} gives $0.9288/q$.

Finally we note that the transverse structure factor exceeds the longitudinal
one throughout the zone. The dominant one-magnon states only contribute to the transverse
structure factor.
%\bea  % 1.55827046D-02  4.0D-04 (t=0)    1.59952368D-02  2.5D-04  (t=1)
%S_l ({\bf k}) &=& 0.042(4)k^2  \quad {\bf k} \to 0  %  2D=  8.417810D-02  6.2D-03
%S_{\rm tot}(k)&=& 0.109(2) k, \quad k\to 0 \\      %v^2= 1.191005D-02  3.3D-04 (d1)
%S_t ({\bf k}) &=& 0.108(4) k,  \quad {\bf k} \to 0 \\ %v^2= 1.169223D-02  7.3D-04 (d1)
%A_{\rm 1} (k) &=& 0.111(5) k , \quad k\to 0 \\ % v^2= 1.252982D-02  9.1D-04 (d1)
%%
%S_{\rm t} (k) &=& 0.93(7)/k , \quad k\to (\pi,\pi,\pi) \\  %v^(-2)=1.15(15) (d1)  1.05(8)(x1)
%S_{\rm tot}(k)&=& 0.95(5)/k , \quad k\to (\pi,\pi,\pi) \\  %v^(-2)=1.1(1)
%A_{\rm 1} (k) &=& 0.97(5)/k , \quad k\to (\pi,\pi,\pi) \\  %v^(-2)=1.05(10)(d1)   1.051D+00  5.5D-03
%\eea
The data can be analysed to extract the 1-magnon spectral weight $A_1({\bf k})$ and
the relative multi-magnon spectral weights (Eqs. \ref{eq_W_t}, \ref{eq_W_tot}).
These are shown in Figure \ref{fig_wei_1p_sq}, for the conventional lines in the Brillouin zone.
The total 1-magnon spectral weight, summed over ${\bf k}$, has the value 0.419(2),
i.e. the 1-magnon excitations contribute $0.419/0.5\simeq 84\%$
%Our calculations shows that the autocorrelation function  for
%the 1-magnon structure factor is 0.419(2),
%i.e. the single magnon carries 0.419/0.5=83.8(4)\%
of the total transverse weight.
We note that the maximum multi-magnon contribution to the structure factors, and
hence to the integrated neutron scattering intensity, occurs at the $(\pi,0)$ point
and is approximately 44\% (29\%)
for unpolarized (polarized) neutrons. For ${\bf k} = (\pi/2,\pi/2)$,
the multi magnon contribution is 31\% (10\%)
for unpolarized (polarized) neutrons.
Quantum Monte Carlo calculations\cite{sandvik} give 40\% (15\%) at
 ${\bf k} = (\pi,0)$ (${\bf k}=(\pi/2,\pi/2)$) for polarized neutrons.
This is a significant contribution and needs to be allowed for in analysis of experimental data.

%To compare our results with other calculations we plot in Figure \ref{fig_S_k0}(a,b)
%the quantities $S_t(k)/k$ and $k S_t(k)$
%along the lines $(0,0)\to (\pi,0)$ and $(0,0)\to (\pi/2,\pi/2)$ (??).
%The results of 1st and 2nd order spin-wave theory\cite{?} are shown for
%comparison. ???
In Figure \ref{fig_ratio_Szz_Sxy_sq} we plot the ratio $2S_l/S_t$ throughout the zone.
The overall shape is in  excellent agreement  with recent Quantum Monte Carlo data\cite{sandvik},
but our maximum is about 0.62, considerably  lower than the value 0.7 obtained by the
Monte Carlo calculations\cite{sandvik}.
Note that the Quantum Monte Carlo calculations have $S_t/S_l$ diverging at
${\bf k} = (\pi,\pi)$, as they do not
include the term (\ref{eq_add_term_lsf})
in their definition of the longitudinal structure factor.
In principle, this term is a simple delta function at $(\pi,\pi)$,
and should not affect the measurement elsewhere for the bulk system.
The omission  of this term in the Monte Carlo calculations, however,
can cause larger finite-size effects for finite systems, and this could be
the cause of the discrepancy.
%We have no explanation at the present
%time for these discrepancies between the series and Monte Carlo data.

\section{\label{sec5}The Simple Cubic Lattice}
We have carried out similar series calculations for the simple cubic
lattice, and report on these here. Firstly, the previously calculated
series for the ground state properties\cite{oit94} have been extended by
two terms, to order $\lambda^{14}$, involving a list of
180\,252 clusters, up to 14 sites. This does not significantly
change the previous estimates of ground state properties, and we do not
present any further analysis. As usual, we are happy to provide
the new coefficients to any interested reader.

\begin{figure}[htb]
\centering
      \includegraphics[width=0.45\textwidth]{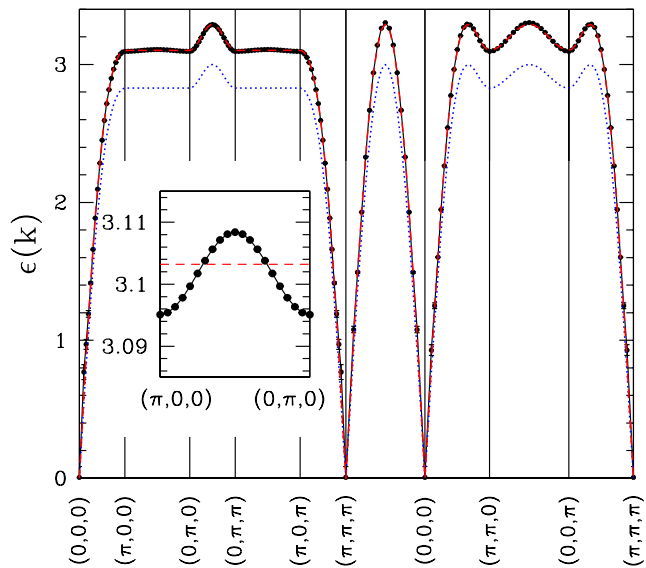}
      \caption{(Color online)  The 1-magnon excitation spectrum
$\epsilon ( {\bf k})$ along high-symmetry cuts through the Brillouin
zone for the Heisenberg antiferromagnet on the simple cubic lattice. Also shown are
      the results of  first order (blue dotted line) and second order (red dashed line)  spin-wave theory.
     \label{fig_mk_sc}}
\end{figure}

Series for the magnon excitation spectrum have been derived, for the first
time, to order $\lambda^{10}$. The calculations involve a list of
1\,487\,597 clusters, up to 11 sites.
The series for ${\bf k}=(\pi,\pi,\pi)$,
$(\pi,0,0)$, $(\pi/2,\pi/2,\pi/2)$ are given in Table \ref{tab_sc}.
 Figure \ref{fig_mk_sc} shows the
magnon excitation spectrum along high-symmetry lines through the Brillouin zone,
obtained from the series expansion, and  first and second order spin-wave theory.
It is evident from the figure that first order spin-wave theory gives
the correct overall shape, but underestimates the magnitude by some 10\%.
The second order spin-wave theory is virtually indistinguishable
from the series data, except on an enlarged scale along some cuts
(as shown in the inset). A calculation of the spin-wave
velocity, along the same lines as in the previous section,
yields  $v/Ja = 1.913(2)$.
This compares with the first (second) order  spin-wave value of $3^{1/2}=1.732$ (1.9003),
and yields a quantum renormalization factor of $Z=1.104(1)$
(compared to the square lattice with $Z=1.178(2)$).
This again is totally consistent with the lower relative effect of quantum
fluctuations in higher dimensions.

\begin{figure}[htb]
\centering
      \includegraphics[width=0.45\textwidth]{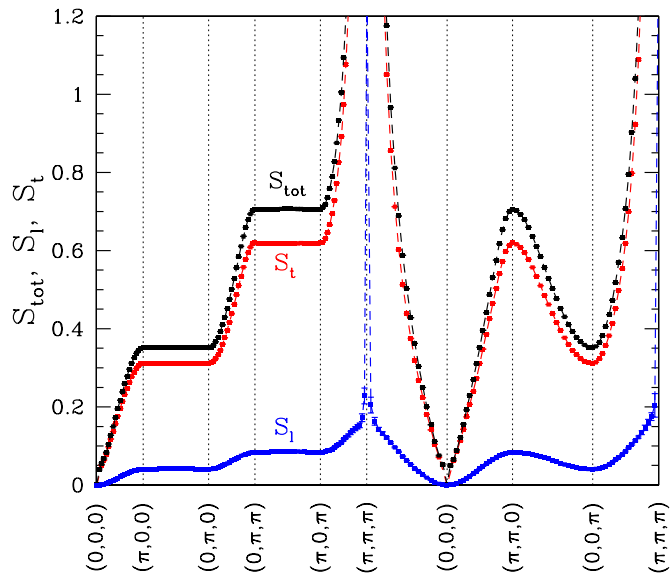}
      \caption{(Color online) The various integrated structure factors $S_{\rm tot}$ (unpolarized)
      , $S_{\rm t}$ (transverse)  and $S_{\rm l}$ (longitudinal) along high-symmetry cuts through the Brillouin
zone for the Heisenberg antiferromagnet on a simple cubic lattice.\label{fig_tot_wei_sc}}
\end{figure}

\begin{figure}[htb]
\centering
      \includegraphics[width=0.45\textwidth]{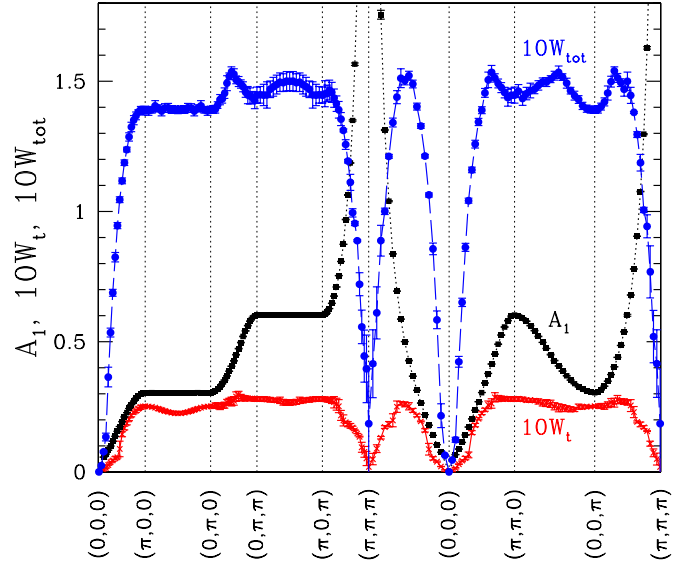}
      \caption{(Color online) The 1-magnon spectral weight $A_{\rm 1}$, and  multi-magnon spectral weights
      $W_{\rm t}$ and $W_{\rm tot}$ for the Heisenberg antiferromagnet on the simple cubic lattice.\label{fig_1p_wei_sc}}
\end{figure}

\begin{figure}[htb]
\centering
      \includegraphics[width=0.45\textwidth]{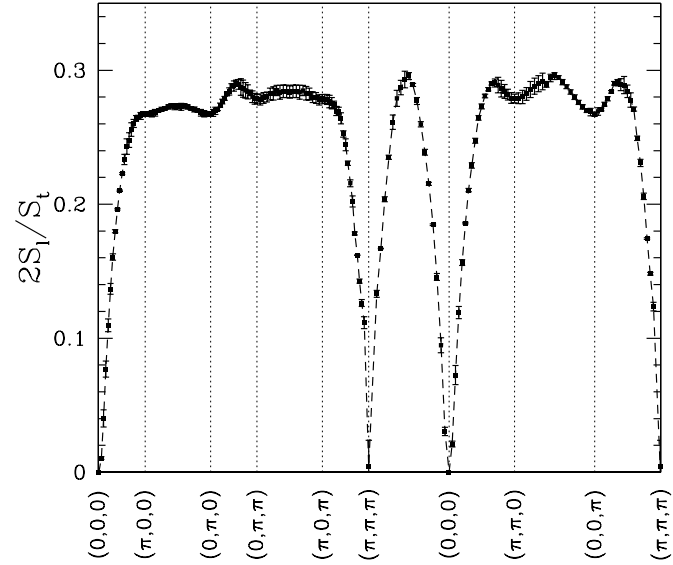}
      \caption{Ratio of longitudinal and transverse structure factors $2S_{\rm l}/S_{\rm t}$
      for the Heisenberg antiferromagnet on the simple cubic lattice.\label{fig_ratio_Szz_Sxy_sc}}
\end{figure}

Figure \ref{fig_tot_wei_sc} gives our series estimates of the integrated
structure factors along symmetry lines in the simple cubic Brillouin zone.
These are obtained from series expansions to order $\lambda^{10}$.
We are unaware of any previous work along these lines.
The same observations made for the square lattice can be made here.
We note that the antiferromagnetic peak in $S_l$ is noticeably sharper
here than for the square lattice, again reflecting the reduced role
of quantum fluctuations. Finally in Figure \ref{fig_1p_wei_sc} we show
the 1-magnon spectral weight and the relative multi-magnon spectral weights.
The latter are magnified by a factor of 10 for greater clarity.
The multimagnon contribution to the transverse structure factor is nowhere greater than
3\%, indicating the dominance of 1-magnon states, while the multimagnon contribution
to the total structure factor, as would be measured by unpolarized neutrons, is as much as 15\%.
% Our calculation shows that  the autocorrelation function  for
% the 1-magnon structure factor is 0.482(1), i.e. one magnon state has 96.4(2)\%
% of the total transverse weight.
The total 1-magnon spectral weight, summed over ${\bf k}$, has the value 0.482(1),
i.e. the 1-magnon excitations contribute  96.4\%
of the total transverse weight.

Similarly to the square lattice case, we obtain  the following asymptotic results near ${\bf k}=0$
and $(\pi,\pi,\pi)$:
% sw v = 3^(1/2),  (1+C1) 3^(1/2) = 1.097158 3^(1/2)=1.90033
\bea
%\epsilon ({\bf k})  &=& 1.913(2)  k  \quad {\bf k}\to 0 \\ %v^2 = 3.65632094D+00  4.1D-03
S_{\rm l} ({\bf k}) &=& 0.0114(2) k^2, \quad {\bf k}\to 0 \\  %2D = 0.0227(4)
%S_{\rm tot}({\bf k})&=& 0.1207(6) k , \quad {\bf k}\to 0 \\ % v^2=1.459543D-02  5.9D-05(d1)  1.776638D-02 1.6D-04(x1) using extrapolated in variable delta=1-(1-x)^(1/2)
S_{\rm t} ({\bf k}) &=& 0.1204(9) k , \quad {\bf k}\to 0 \\ % v^2=1.450436D-02  9.3D-05(d1)  1.614845D-02 1.3D-04(x1)
%A_{\rm 1} (k) &=& 0.1203(8) k , \quad {\bf k}\to 0 \\ % v^2=1.44755928D-02  7.1D-04(d1)
S_{\rm t} ({\bf q}) &=& 1.47(3)/q , \quad {\bf q}\to 0   %v^(-2)=4.80D-01 2.5D-03(d1)  4.411D-01 4.D-03(x1) using extrapolated in variable delta=1-(1-x)^(1/2)
%S_{\rm tot}({\bf q})&=& 1.50(5)/q , \quad {\bf q}\to 0   %v^(-2)=4.42D-01 1.6D-02(d1)  3.692D-01 8.9D-03(x1)
%A_{\rm 1} ({\bf q}) &=& 1.45(5)/q , \quad {\bf q}\to 0   %v^(-2)=4.329D-01  6.3D-03(d1)  using extrapolated in variable delta=1-(1-x)^(1/2)
\eea
Estimates from the $S_{\rm tot}$ series are consistent with these.

In Figure \ref{fig_ratio_Szz_Sxy_sc} we plot the ratio $2S_l/S_t$ throughout the zone.
Here it has a maximum value about 0.3, substantially smaller than for the square lattice.
We are unaware of any calculations of this ratio by other methods.

\begin{figure}[htb]
\centering
      \includegraphics[width=0.45\textwidth]{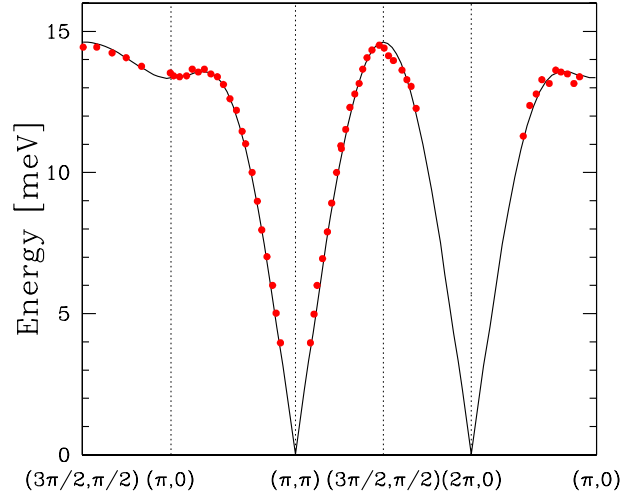} % fig10_CFTD_fit_J6.34.ps}
      \caption{(Color online) Comparison of the 1-magnon dispersion for CFTD\cite{Ronnow}  (red solid points) and
      our series results with $J=6.13 {\rm meV}$. %  $J=6.34 {\rm meV}$.
     \label{fig_CFTD}}
\end{figure}

\begin{figure}[htb]
\centering
      \includegraphics[width=0.45\textwidth]{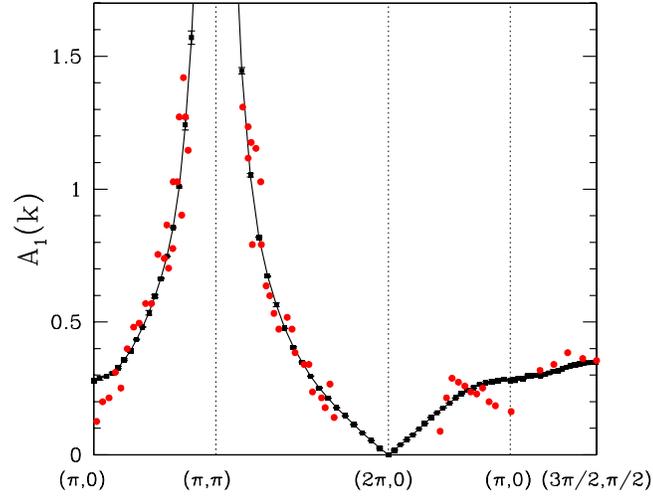}
      \caption{(Color online) Comparison of the 1-magnon transverse structure factor $A_1({\bf k})$  for CFTD\cite{Ronnow}
      (red solid points)
      and our series results.
     \label{fig_CFTD2}}
\end{figure}

\begin{figure}[htb]
\centering
      \includegraphics[width=0.45\textwidth]{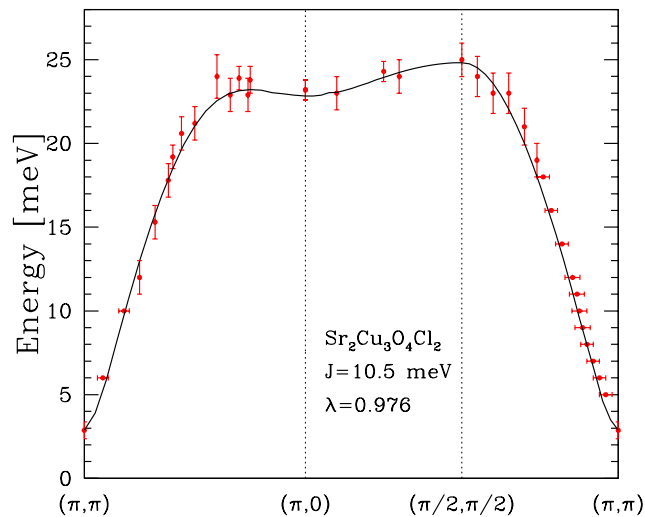}
      \caption{Comparison of the 1-magnon dispersion for Sr$_2$Cu$_3$O$_4$Cl$_2$\cite{Kim} and our
      series results with $J=10.5 {\rm meV}$, $\lambda=0.976$.
     \label{fig_SrCuOCl}}
\end{figure}

\section{\label{sec6}Summary and Discussions}

The goal of this work has been to obtain numerically precise estimates of magnon energies and
structure factors for the nearest-neighbour spin-$\half$ Heisenberg antiferromagnet for the linear chain
($d=1$), square ($d=2$) and simple cubic ($d=3$) lattices. These quantities are directly comparable to
experimental neutron scattering results, and the resulting comparison can provide a stringest test
of the applicability of the simple model, as well as yielding an estimate of the (usually unknown)
parameter $J$.

We present such a comparison here for the quasi two-dimensional
materials deuterated copper formate tetrahydrate  (CuDCOO)$_2\cdot 4$D$_2$O (CFTD)\cite{Ronnow}
%copper deuteroformate tetradeuterate
and the so-called ``2342" compound
Sr$_2$Cu$_3$O$_4$Cl$_2$.\cite{Kim}
CFTD is a well characterized 2-d antiferromagnet\cite{Ronnow}. Figures \ref{fig_CFTD} and \ref{fig_CFTD2}
show a fit of our theoretical dispersion curve (Fig. \ref{fig_mk_sq}) and 1-magnon transverse structure
factor $A_1({\bf k})$ to the experimental data\cite{Ronnow}, with a parameter
$J=6.13 {\rm meV}$.
% $J=6.34 {\rm meV}$.
The overall agreement is very good, except near ${\bf k}=(\pi,0)$, where the theoretical
one-magnon transverse structure factor is higher than the experimental results.
The fitting parameter $J$ is in %excellent
good agreement with an earlier fit\cite{Ronnow}
to the previous series results\cite{rajiv}.
 The  strontium  material is, a priori, more complex\cite{Kim}.
It contains two types of Cu$^{2+}$ ions, Cu$_{\rm I}$ and Cu$_{\rm II}$,
and the interaction between these is fully frustrated. To the extent that one can regard
these subsystems as decoupled, the   Cu$_{\rm II}$ subsystem can be treated as an effective
spin-$\half$ square lattice antiferromagnet with $J\sim 10 {\rm meV}$.
The measured dispersion curve shows a small spin gap, which can be modelled via
a small magnetic anisotropy in the Hamiltonian. Figure \ref{fig_SrCuOCl} shows
a comparison between the experimental data and our series results with $J=10.5 {\rm meV}$,
$\lambda=0.976$, where $\lambda$ is determined from the  minimum gap using Eq. (\ref{eq_gap_k0}).
As is evident the fit is excellent, and again corroborates
earlier results\cite{Kim}. One should be cautious, however, in claiming too much from this and it would
be highly desirable to have detailed structure factor data for further comparisons to be made.

We are unaware of any good examples of spin-$\half$ antiferromagnetic
materials with a simple cubic structure.

Our results confirm, as expected, that the relative effect of quantum fluctuations
decreases with increasing spatial dimension. Nevertheless, the multi-magnon
contributions to integrated structure factors, and hence to neutron scattering
intensities, can still be appreciable even in three dimensions.
%Our results clearly show that the multi-magnon contributions to integrated structure factors, and
%hence to neutron scattering intensities, are not insignificant and need to be allowed for in
%any analysis of experimental data. Our results also confirm, as expected, that the relative effect
%of quantum fluctuations decreases with increasing spatial dimension.

For dimensions 2 and 3, the series expansion results are in very good agreement with spin-wave
theory, as far as it has been calculated. We conclude that the spin-wave calculations should be extended to higher
order, to further check the agreement in quantities, such as the longitudinal structure factor, which
have been little studied as yet.
% The third order spin-wave theory
% has already been calculated for the minimum gap\cite{zwh_sw}, but not for the
% full magnon spectrum. It would be interesting to see whether the spectrum remains dispersionless along
% the line $(\pi,0)$ to $(\pi/2,\pi/2)$ for the square lattice.
%existed for the minimum spin-wave
%energy gap\cite{zwh_sw}, one should also extend this calculation to the full spin-wave dispersion to see
%whether it can give the dispersion along the magnetic Brillouin zone
%boundary for square lattice.

Note added: after this paper was submitted, we became aware of the
work by Igarashi and  Nagao\cite{Igarashi}, who have performed a
second-order spin-wave calculation of the transverse
 structure factor for the square lattice.
\begin{acknowledgments}
We are grateful to Professors Rajiv Singh, T. Barnes,  R. Coldea, and A. Sandvik for
useful correspondences. This work  forms part of a research project supported by
a grant from the Australian Research Council.
We are grateful for the computing resources provided
by the Australian Partnership for Advanced Computing (APAC)
National Facility and by the
Australian Centre for Advanced Computing and Communications (AC3).
\end{acknowledgments}

\newpage

\bibliography{paper.bst}

% \newpage

%\setdec 0.000000000000
\begin{table*}
\squeezetable
\caption{Series of square lattice one-magnon dispersion $\epsilon({\bf k})$,
longitudinal structure factor  $S_{\rm l} ({\bf k})$,
transverse structure factor  $S_{\rm t} ({\bf k})$,
and one-magnon exclusive structure factor  $A_{1} ({\bf k})$
at ${\bf k}=(\pi,\pi)$,
$(\pi,0)$, $(\pi/2,\pi/2)$, and series $D$ for coefficient
of $k^2$ (for $\epsilon$ and $S_{\rm l} $ )
or $q^2$ (for $S_{\rm t} $ and $A_{1} $).
Nonzero coefficients $\lambda^n$ up to order $n=14$  are listed. }\label{tab_sq}
\begin{ruledtabular}
\begin{tabular}{|r|l|l|l|l|}
\multicolumn{1}{|c|}{$n$} & \multicolumn{1}{c|}{${\bf k}=(\pi,\pi)$} & \multicolumn{1}{c|}{${\bf k}=(\pi,0)$}
& \multicolumn{1}{c|}{${\bf k}=(\pi/2,\pi/2)$} & \multicolumn{1}{c|}{$D$} \\
\hline
\multicolumn{5}{|c|}{dispersion  $\epsilon({\bf k})$}\\
  0 & ~~2.000000000     & ~~2.000000000     & ~~2.000000000     & ~~0.000000000     \\
  2 & $-$1.666666667     & ~~3.333333333$\times 10^{-1}$ & ~~3.333333333$\times 10^{-1}$ & ~~1.000000000     \\
  4 & ~~3.171296296$\times 10^{-1}$ & $-$9.953703704$\times 10^{-2}$ & ~~5.324074074$\times 10^{-2}$ & ~~2.569444444$\times 10^{-1}$ \\
  6 & $-$4.192337641$\times 10^{-1}$ & $-$1.693897891$\times 10^{-3}$ & $-$9.073302469$\times 10^{-3}$ & ~~6.581836259$\times 10^{-1}$ \\
  8 & ~~2.709969904$\times 10^{-1}$ & $-$2.806720342$\times 10^{-2}$ & ~~5.105325304$\times 10^{-3}$ & $-$4.886280904$\times 10^{-2}$ \\
 10 & $-$3.894335149$\times 10^{-1}$ & $-$1.062177000$\times 10^{-2}$ & ~~2.076320167$\times 10^{-3}$ & ~~7.984148319$\times 10^{-1}$ \\
 12 & ~~4.289652578$\times 10^{-1}$ & $-$9.046042545$\times 10^{-3}$ & ~~4.068538933$\times 10^{-4}$ & $-$5.057247719$\times 10^{-1}$ \\
 14 & $-$6.558882026$\times 10^{-1}$ & $-$8.872458632$\times 10^{-3}$ & ~~1.304340621$\times 10^{-3}$ & ~~1.460095528     \\
\hline
\multicolumn{5}{|c|}{longitudinal structure factor  $S_{\rm l} ({\bf k})$}\\
%  0 &  ~~0.000000000     & ~~0.000000000     & ~~0.000000000     & ~~0.000000000     \\
  2 &  ~~2.222222222$\times 10^{-1}$ & ~~1.111111111$\times 10^{-1}$ & ~~1.111111111$\times 10^{-1}$ & ~~2.777777778$\times 10^{-2}$ \\
  4 &  ~~7.358024691$\times 10^{-2}$ & $-$1.580246914$\times 10^{-2}$ & ~~3.703703704$\times 10^{-4}$ & ~~1.851851852$\times 10^{-3}$ \\
  6 &  ~~4.055166849$\times 10^{-2}$ & ~~1.167542552$\times 10^{-2}$ & $-$1.111176199$\times 10^{-3}$ & ~~3.826627719$\times 10^{-3}$ \\
  8 &  ~~3.646524757$\times 10^{-2}$ & ~~1.065352379$\times 10^{-3}$ & ~~4.750603835$\times 10^{-3}$ & ~~2.311003576$\times 10^{-3}$ \\
 10 &  ~~2.483688972$\times 10^{-2}$ & ~~2.298840789$\times 10^{-3}$ & ~~1.379013443$\times 10^{-3}$ & ~~1.874238376$\times 10^{-3}$ \\
 12 &  ~~2.211921434$\times 10^{-2}$ & ~~1.532346963$\times 10^{-3}$ & ~~1.422483566$\times 10^{-3}$ & ~~1.550055137$\times 10^{-3}$ \\
 14 &  ~~1.886279612$\times 10^{-2}$ & ~~1.141259073$\times 10^{-3}$ & ~~1.002657611$\times 10^{-3}$ & ~~1.314140042$\times 10^{-3}$ \\
\hline
\multicolumn{5}{|c|}{transverse structure factor  $S_{\rm t} ({\bf k})$}\\
  0 & ~~5.000000000$\times 10^{-1}$ & ~~5.000000000$\times 10^{-1}$ & ~~5.000000000$\times 10^{-1}$ & ~~0.000000000     \\
  1 & ~~6.666666667$\times 10^{-1}$ & ~~0.000000000     & ~~0.000000000     & $-$1.666666667$\times 10^{-1}$ \\
  2 & ~~3.333333333$\times 10^{-1}$ & $-$1.111111111$\times 10^{-1}$ & $-$1.111111111$\times 10^{-1}$ & $-$2.222222222$\times 10^{-1}$ \\
  3 & ~~2.648148148$\times 10^{-1}$ & ~~0.000000000     & ~~0.000000000     & $-$3.773148148$\times 10^{-1}$ \\
  4 & ~~2.390123457$\times 10^{-1}$ & ~~1.925925926$\times 10^{-2}$ & ~~1.086419753$\times 10^{-2}$ & $-$4.340740741$\times 10^{-1}$ \\
  5 & ~~2.157488242$\times 10^{-1}$ & ~~0.000000000     & ~~0.000000000     & $-$5.523841490$\times 10^{-1}$ \\
  6 & ~~1.922286407$\times 10^{-1}$ & $-$1.297374724$\times 10^{-2}$ & $-$6.735355253$\times 10^{-3}$ & $-$6.036591991$\times 10^{-1}$ \\
  7 & ~~1.825523316$\times 10^{-1}$ & ~~0.000000000     & ~~0.000000000     & $-$7.022537598$\times 10^{-1}$ \\
  8 & ~~1.672647956$\times 10^{-1}$ & $-$2.593884626$\times 10^{-4}$ & $-$2.653241665$\times 10^{-3}$ & $-$7.488016779$\times 10^{-1}$ \\
  9 & ~~1.584816304$\times 10^{-1}$ & ~~0.000000000     & ~~0.000000000     & $-$8.361027757$\times 10^{-1}$ \\
 10 & ~~1.488115597$\times 10^{-1}$ & $-$2.819695062$\times 10^{-3}$ & $-$1.579465770$\times 10^{-3}$ & $-$8.786542283$\times 10^{-1}$ \\
 11 & ~~1.436901069$\times 10^{-1}$ & ~~0.000000000     & ~~0.000000000     & $-$9.576556162$\times 10^{-1}$ \\
 12 & ~~1.358134394$\times 10^{-1}$ & $-$1.233261351$\times 10^{-3}$ & $-$1.422867383$\times 10^{-3}$ & $-$9.970882006$\times 10^{-1}$ \\
 13 & ~~1.316662910$\times 10^{-1}$ & ~~0.000000000     & ~~0.000000000     & $-$1.069785636     \\
 14 & ~~1.256571308$\times 10^{-1}$ & $-$1.266236691$\times 10^{-3}$ & $-$1.061011294$\times 10^{-3}$ & $-$1.106759542     \\
\hline
\multicolumn{5}{|c|}{one-magnon spectral weight  $A_{1} ({\bf k})$}\\
  0 & ~~5.000000000$\times 10^{-1}$ & ~~5.000000000$\times 10^{-1}$ & ~~5.000000000$\times 10^{-1}$ & ~~0.000000000     \\
  1 & ~~6.666666667$\times 10^{-1}$ & ~~0.000000000     & ~~0.000000000     & $-$1.666666667$\times 10^{-1}$ \\
  2 & ~~2.500000000$\times 10^{-1}$ & $-$1.388888889$\times 10^{-1}$ & $-$1.388888889$\times 10^{-1}$ & $-$1.944444444$\times 10^{-1}$ \\
  3 & ~~1.425925926$\times 10^{-1}$ & ~~0.000000000     & ~~0.000000000     & $-$3.245370370$\times 10^{-1}$ \\
  4 & ~~3.326195988$\times 10^{-1}$ & $-$1.143904321$\times 10^{-2}$ & ~~1.529706790$\times 10^{-2}$ & $-$5.072723765$\times 10^{-1}$ \\
  5 & ~~3.917638154$\times 10^{-1}$ & ~~0.000000000     & ~~0.000000000     & $-$6.858476264$\times 10^{-1}$ \\
  6 & ~~3.936459588$\times 10^{-2}$ & $-$1.917201533$\times 10^{-2}$ & $-$1.696167024$\times 10^{-2}$ & $-$4.528242195$\times 10^{-1}$ \\
  7 & $-$8.018312217$\times 10^{-2}$ & ~~0.000000000     & ~~0.000000000     & $-$4.216442748$\times 10^{-1}$ \\
  8 & ~~3.778813233$\times 10^{-1}$ & $-$1.855816875$\times 10^{-2}$ & $-$4.208337928$\times 10^{-3}$ & $-$1.027026315     \\
  9 & ~~5.220328525$\times 10^{-1}$ & ~~0.000000000     & ~~0.000000000     & $-$1.343102568     \\
 10 & $-$1.799044627$\times 10^{-1}$ & $-$1.153727416$\times 10^{-2}$ & $-$3.162792993$\times 10^{-3}$ & $-$3.701268973$\times 10^{-1}$ \\
 11 & $-$4.037813448$\times 10^{-1}$ & ~~0.000000000     & ~~0.000000000     & $-$4.716188047$\times 10^{-2}$ \\
 12 & ~~6.539878806$\times 10^{-1}$ & $-$9.013887872$\times 10^{-3}$ & $-$3.821489172$\times 10^{-3}$ & $-$1.930735417     \\
 13 & ~~9.853540532$\times 10^{-1}$ & ~~0.000000000     & ~~0.000000000     & $-$2.711397791     \\
 14 & $-$7.366801888$\times 10^{-1}$ & $-$9.184484559$\times 10^{-3}$ & $-$1.633241155$\times 10^{-3}$ & ~~6.457723508$\times 10^{-1}$ \\
\end{tabular}
\end{ruledtabular}
\end{table*}

%\setdec 0.000000000000
\begin{table*}
\squeezetable
\caption{Series of simple cubic lattice  one-magnon dispersion $\epsilon({\bf k})$,
longitudinal structure factor  $S_{\rm l} ({\bf k})$,
transverse structure factor  $S_{\rm t} ({\bf k})$,
and one-magnon exclusive structure factor  $A_{1} ({\bf k})$
at ${\bf k}=(\pi,\pi,\pi)$,
$(\pi,0,0)$, $(\pi/2,\pi/2,\pi/2)$, and series $D$ for coefficient
of $k^2$ (for $\epsilon$ and $S_{\rm l} $ )
or $q^2$ (for $S_{\rm t} $ and $A_{1} $).
Nonzero coefficients $\lambda^n$ up to order $n=10$  are listed. }\label{tab_sc}
\begin{ruledtabular}
\begin{tabular}{|r|l|l|l|l|}
\multicolumn{1}{|c|}{$n$} & \multicolumn{1}{c|}{${\bf k}=(\pi,\pi,\pi)$} & \multicolumn{1}{c|}{${\bf k}=(\pi,0,0)$}
& \multicolumn{1}{c|}{${\bf k}=(\pi/2,\pi/2,\pi/2)$} & \multicolumn{1}{c|}{$D$} \\
\hline
\multicolumn{5}{|c|}{dispersion  $\epsilon({\bf k})$}\\
  0 &  ~~3.000000000     &  ~~3.000000000     &  ~~3.000000000     &  ~~0.000000000     \\
  2 & $-$1.950000000     &  ~~5.000000000$\times 10^{-2}$ &  ~~3.000000000$\times 10^{-1}$ &  ~~7.500000000$\times 10^{-1}$ \\
  4 & $-$7.480952381$\times 10^{-2}$ &  ~~5.483333333$\times 10^{-2}$ & $-$1.550595238$\times 10^{-3}$ &  ~~2.859821429$\times 10^{-1}$ \\
  6 & $-$2.386949857$\times 10^{-1}$ & $-$1.194666672$\times 10^{-2}$ &  ~~3.986473230$\times 10^{-3}$ &  ~~3.080615484$\times 10^{-1}$ \\
  8 & $-$3.884790029$\times 10^{-2}$ &  ~~2.654093113$\times 10^{-3}$ &  ~~6.579308964$\times 10^{-4}$ &  ~~1.842288801$\times 10^{-1}$ \\
 10 & $-$1.028725933$\times 10^{-1}$ & $-$5.327445001$\times 10^{-4}$ &  ~~5.131262221$\times 10^{-4}$ &  ~~2.278014891$\times 10^{-1}$ \\
\hline
\multicolumn{5}{|c|}{longitudinal structure factor  $S_{\rm l} ({\bf k})$}\\
  2 &  ~~1.200000000$\times 10^{-1}$ &  ~~4.000000000$\times 10^{-2}$ &  ~~6.000000000$\times 10^{-2}$ &  ~~1.000000000$\times 10^{-2}$ \\
  4 &  ~~1.684444444$\times 10^{-2}$ & $-$2.488888889$\times 10^{-3}$ & $-$3.962962963$\times 10^{-4}$ &  ~~1.530864198$\times 10^{-4}$ \\
  6 &  ~~1.301036907$\times 10^{-2}$ &  ~~2.757538780$\times 10^{-3}$ &  ~~2.815062245$\times 10^{-3}$ &  ~~8.063018305$\times 10^{-4}$ \\
  8 &  ~~8.184237447$\times 10^{-3}$ &  ~~7.116986056$\times 10^{-4}$ &  ~~1.126836657$\times 10^{-3}$ &  ~~3.866267816$\times 10^{-4}$ \\
 10 &  ~~6.005783346$\times 10^{-3}$ &  ~~6.521285585$\times 10^{-4}$ &  ~~8.514828105$\times 10^{-4}$ &  ~~3.009882557$\times 10^{-4}$ \\
\hline
\multicolumn{5}{|c|}{transverse structure factor  $S_{\rm t} ({\bf k})$}\\
  0 &  ~~5.000000000$\times 10^{-1}$ &  ~~5.000000000$\times 10^{-1}$ &  ~~5.000000000$\times 10^{-1}$ &  ~~0.000000000     \\
  1 &  ~~6.000000000$\times 10^{-1}$ & $-$2.000000000$\times 10^{-1}$ &  ~~0.000000000     & $-$1.000000000$\times 10^{-1}$ \\
  2 &  ~~3.000000000$\times 10^{-1}$ & $-$2.000000000$\times 10^{-2}$ & $-$6.000000000$\times 10^{-2}$ & $-$1.200000000$\times 10^{-1}$ \\
  3 &  ~~2.526666667$\times 10^{-1}$ &  ~~4.614814815$\times 10^{-2}$ &  ~~0.000000000     & $-$1.887777778$\times 10^{-1}$ \\
  4 &  ~~2.137481481$\times 10^{-1}$ & $-$1.191111111$\times 10^{-2}$ &  ~~3.644444444$\times 10^{-3}$ & $-$2.033679012$\times 10^{-1}$ \\
  5 &  ~~2.025150853$\times 10^{-1}$ & $-$2.558718236$\times 10^{-3}$ &  ~~0.000000000     & $-$2.520957812$\times 10^{-1}$ \\
  6 &  ~~1.752635491$\times 10^{-1}$ & $-$7.923299511$\times 10^{-4}$ & $-$3.718193643$\times 10^{-3}$ & $-$2.638276824$\times 10^{-1}$ \\
  7 &  ~~1.685481230$\times 10^{-1}$ &  ~~9.547987414$\times 10^{-4}$ &  ~~0.000000000     & $-$3.039691386$\times 10^{-1}$ \\
  8 &  ~~1.523090399$\times 10^{-1}$ & $-$1.590182006$\times 10^{-3}$ & $-$9.298721272$\times 10^{-4}$ & $-$3.138723809$\times 10^{-1}$ \\
  9 &  ~~1.480721342$\times 10^{-1}$ &  ~~5.897338735$\times 10^{-4}$ &  ~~0.000000000     & $-$3.487614896$\times 10^{-1}$ \\
 10 &  ~~1.365201003$\times 10^{-1}$ & $-$9.432640534$\times 10^{-4}$ & $-$8.927606509$\times 10^{-4}$ & $-$3.574938809$\times 10^{-1}$ \\
\hline
\multicolumn{5}{|c|}{one-magnon spectral weight  $A_{1} ({\bf k})$}\\
  0 &  ~~5.000000000$\times 10^{-1}$ &  ~~5.000000000$\times 10^{-1}$ &  ~~5.000000000$\times 10^{-1}$ &  ~~0.000000000     \\
  1 &  ~~6.000000000$\times 10^{-1}$ & $-$2.000000000$\times 10^{-1}$ &  ~~0.000000000     & $-$1.000000000$\times 10^{-1}$ \\
  2 &  ~~2.812500000$\times 10^{-1}$ & $-$2.875000000$\times 10^{-2}$ & $-$6.750000000$\times 10^{-2}$ & $-$1.162500000$\times 10^{-1}$ \\
  3 &  ~~2.251666667$\times 10^{-1}$ &  ~~5.309259259$\times 10^{-2}$ &  ~~0.000000000     & $-$1.816944444$\times 10^{-1}$ \\
  4 &  ~~2.268800324$\times 10^{-1}$ & $-$1.446118552$\times 10^{-2}$ &  ~~1.009873984$\times 10^{-3}$ & $-$2.105439590$\times 10^{-1}$ \\
  5 &  ~~2.300857299$\times 10^{-1}$ & $-$5.633309085$\times 10^{-3}$ &  ~~0.000000000     & $-$2.648295019$\times 10^{-1}$ \\
  6 &  ~~1.606446880$\times 10^{-1}$ & $-$2.164942522$\times 10^{-3}$ & $-$4.497506567$\times 10^{-3}$ & $-$2.550269987$\times 10^{-1}$ \\
  7 &  ~~1.424582061$\times 10^{-1}$ &  ~~2.355649892$\times 10^{-3}$ &  ~~0.000000000     & $-$2.873096589$\times 10^{-1}$ \\
  8 &  ~~1.648269861$\times 10^{-1}$ & $-$2.195243132$\times 10^{-3}$ & $-$1.493247410$\times 10^{-3}$ & $-$3.243370831$\times 10^{-1}$ \\
  9 &  ~~1.707585298$\times 10^{-1}$ &  ~~2.315207796$\times 10^{-4}$ &  ~~0.000000000     & $-$3.678986537$\times 10^{-1}$ \\
 10 &  ~~1.243609563$\times 10^{-1}$ & $-$1.435173150$\times 10^{-3}$ & $-$1.241136381$\times 10^{-3}$ & $-$3.458041669$\times 10^{-1}$ \\
 \end{tabular}
\end{ruledtabular}
\end{table*}

\end{document}